\documentclass[12pt]{article}
\usepackage{amsmath,amssymb,graphicx,mathrsfs,hyperref,slashed}
\newcommand{\be}{\begin{equation}}
\newcommand{\ee}{\end{equation}}
\newcommand{\bea}{\begin{eqnarray}}
\newcommand{\eea}{\end{eqnarray}}

\newcommand{\wt}{\widetilde}

\def\({\left(} \def\){\right)}

\renewcommand{\baselinestretch}{1.5}
\begin{document}
\title{\vspace{-1.8in}
{ How Black Holes Burn }}
\author{\large Ram Brustein${}^{(1)}$,  A.J.M. Medved${}^{(2)}$ \\
\vspace{-.5in} \hspace{-1.5in} \vbox{
 \begin{flushleft}
  $^{\textrm{\normalsize
(1)\ Department of Physics, Ben-Gurion University,
    Beer-Sheva 84105, Israel}}$
$^{\textrm{\normalsize (2)  Department of Physics \& Electronics, Rhodes University,
  Grahamstown 6140, South Africa }}$
 \\ \small \hspace{1.7in}
    ramyb@bgu.ac.il,\  j.medved@ru.ac.za
\end{flushleft}
}}
\date{}
\maketitle
\begin{abstract}

We present a calculation of the rate of information release from  a Schwarzschild  BH. We have recently extended Hawking's theory of black hole (BH) evaporation to account for quantum fluctuations of the background geometry, as well as for back-reaction and time-dependence effects. Our main result has been a two-point function matrix for the radiation that consists of Hawking's thermal matrix plus off-diagonal corrections that are initially small and become more important as the evaporation proceeds.   Here, we show that, if the phases and amplitudes of the radiation matrix are recorded over the lifetime of the BH, then the radiation purifies in a continuous way.  We conjecture that our results establish the maximal rate at which information can be released from a semiclassical BH, to be contrasted with the minimal rate that was predicted by Page on the basis of generic unitarity arguments. When the phases of the radiation  matrix are not tracked, we show that it purifies only parametrically close to the end of the BH evaporation and does so extremely fast. Our main technical tool in the quantitative treatment of this purification is the purity of the radiation matrix and, its inverse, the participation ratio. These can be related to the Renyi entropy of the density matrix of the emitted radiation.
\end{abstract}

\newpage
\renewcommand{\baselinestretch}{1.5}\normalsize

\section{Introduction}

The tensions between gravitational and quantum theory are pushed to the forefront in Hawking's famed calculation showing that a black hole (BH) emits radiation \cite{Hawk}. On one hand, this is an elegant picture where ideas from gravity, quantum theory and thermodynamics converge in a single arena \cite{Bek}. On the other, accepting the Hawking calculation at face value, one is confronted with the choice of surrendering either quantum unitarity \cite{info} (also, \cite{info2,info3,info4}) or Einstein's equivalence principle \cite{AMPS} (also, \cite{Sunny1,info4,Braun,bousso2}).

Keeping unitarity as the basic principle, Page \cite{page} has proposed the following framework for understanding the phenomenon of BH radiation: Since the BH and the emitted radiation together form a single pure state, he argued that the form of the  density matrix of the Hawking radiation is determined by its own dimensionality relative to that of the BH Hilbert space. He then views the BH as the ``purifier" of the radiation in some random basis, and so  a random unitary transformation relates this basis  to that in which the whole density matrix has a single entry. That such a basis exists is the  minimal requirement that any unitary model of BH evaporation  has to obey. We expect a dynamical model to provide additional information about the nature of this basis.  Page's model thus  defines the minimal amount of information that has to be released from a BH if the evaporation process is unitary and if one has access to all the entries of the radiation density matrix.

The challenge for anyone with a  concrete physical model of BH evaporation is to calculate the amount of information released from the BH and show that the radiation then purifies at a rate that is at least the Page rate. Indeed, Strominger has  already proposed that calculating the rate of information flow from  an evaporating BH  would signify a true understanding of the dynamics and thus a resolution of the BH information paradox \cite{strom1}.

Here,  we present a calculation of the rate of information release from  a Schwarzschild  BH. We calculate the participation ratio ---  the inverse of the purity --- of the two-point function matrix for the Hawking radiation.~\footnote{This matrix was called the ``density matrix'' in our previous articles. We will refer to it here as the ``radiation matrix''
and the ``density matrix for the radiation'' will mean just that.}
The participation ratio can be related to the Renyi entropy of the density matrix of the emitted radiation, as both quantities estimate the number of non-vanishing eigenvalues of the density matrix. The Renyi entropy can be  compared to  the Renyi entropy  in the Page model, thus allowing us to compare the rate of infomation release in our model to that of Page. What we find is that, if the phases and amplitudes of the radiation matrix are recorded over the lifetime of the BH, then the radiation purifies in a continuous way and faster than the Page rate.  On the other hand, when the phases of the radiation matrix are not ``tracked'' in this manner, the radiation purifies at a  slower rate than that of Page's  model. The reason for this discrepancy  is clear: Some of the correlations among the outgoing radiation are missed. The radiation starts to catch up with the Page model only at a time parametrically close to the end of the BH evaporation and does so extremely fast.

It has been suggested by one of the authors that the issues with Hawking's calculation and, more generally, those with  quantum field theory in curved space can be traced to neglecting the fluctuations of the background geometry \cite{RB}. (For similar ideas, see \cite{RJ,Dvali1,RM,Dvali2}.) The importance of accounting for a quantum fluctuating background may be contrary to expectations but has been  explicitly shown to revise some  of the standard premises  of BH physics \cite{flucyou}.

We have, more recently, put this idea to the test by directly addressing  Hawking's model of BH evaporation. Assigning the incipient BH a quantum wavefunction and then repeating the steps of Hawking's calculation, we have found that the radiation matrix picks up perturbatively small corrections \cite{slowleak}. These corrections lead to off-diagonal matrix elements in Hawking's basis that appear negligibly small but act collectively in such a way that purification of the radiation seems feasible. In our first attempt, however, we were only able to make quantitative statements about what happens until the midpoint of evaporation, the so-called Page time \cite{page}.

Later on, we further refined the Hawking calculation  by accounting for, in addition to the background fluctuations,    the different emission times of the particles and how these modes back-react on the collapsing matter \cite{slowburn}.  The main result of the calculation was an explicit form of the radiation  matrix. We have also established  the existence of a novel time scale, the coherence time $t_{coh}$. This scale originates as the time interval over which the BH wavefunction changes significantly and determines the temporal extent of coherent radiation.  This is the time period after emission that the  phases of the emitted  particles are still accessible  without ``tracking'' (in the sense described above). Consequently, $t_{coh}$ sets the rate at which information is transferred out of the BH.

 Typically, the coherence time goes as the square root of the BH entropy in dimensionless units ({\em i.e.}, in units of number of emitted particles) or as the product of the BH entropy and the Planck length in Schwarzschild units. This is parametrically  shorter than the Page time, which scales as the entropy or as the product of the entropy and the Schwarzschild radius, respectively.

The coherence time also  plays a prominent role in defining the phases of information release. For instance, one interval of coherence time after the birth of the BH is when the first bit of information in released. If the radiation amplitude and phases are not tracked (see below) then one interval before the end of evaporation  is when information begins to rapidly stream out. In general, we refer to  the onset of this era of rapid information release as the transparency time $t_{trans}$, although the technical definition of $t_{trans}$ will be slightly modified in the analysis to follow.

In a related discussion \cite{flameoff,noburn}, we have similarly extended Hawking's pair-production picture of BH evaporation \cite{info}. This enabled us to construct a  coefficients matrix for the  state of the in--out pairs that is formally similar to  the radiation  matrix  but having a different interpretation. The off-diagonal elements now represent deviations from the standard thermofield-double state and the coherence  time represents the effective lifetime of a negative energy in-mode.

The remainder of this section reviews the relevant formalism; then
the main analysis and results are  presented  in Section~2.
We  conclude in Section~3 by discussing our results from the pair-production
perspective and conjecturing that our model (with tracking)
establishes the maximal rate of information transfer from a semiclassical BH.

Let us  recall our  multi-particle  two point function matrix for the radiation,
\bea
&&\rho_{SC}(\omega,\wt{\omega};N_T;N',N'') = \frac{1}{N_T}
\rho_H(\omega,\wt{\omega}) \delta_{N',N''}|N'\rangle\langle N''|
\label{nmatrix3} \\
&+&
\frac{C^{1/2}_{BH}(N_T)}{N_T}\Delta\rho_{OD}(\omega,\wt{\omega})\;D(N_T;N',N'')
\ e^{i\Theta_{N' N''}}\ \left[1-\delta_{N',N''}\right]|N'\rangle\langle N''|\;.\nonumber
\eea
Here,  $N_T$ is  the  number of  emitted particles and serves as the time
parameter, $N'$ and $N''$  are particle emission ``times'' ($1\leq N'$,$N''\leq N_T$).
Also, $\;C_{BH}(N_T)=1/S_{BH}(N_T)\;$  is a perturbative ``classicality''
 parameter
and $D(N_T;N',N'')$ is a ``suppression'' factor that is  defined by
\be
D(N_T;N^\prime,N^{\prime\prime})=\frac{1}{2}\left( e^{-\frac{1}{4} \frac{\left[C_{BH}(N^\prime) (N_T-N^\prime)\right]^2}{ C_{BH}(N_T)}}+e^{-\frac{1}{4}\frac{\left[C_{BH}(N^{\prime\prime}) (N_T-N^{\prime\prime})\right]^2}{ C_{BH}(N_T)}}\right)\;.
\label{suppfac}
\ee

The  radiation matrix in Eq.~(\ref{nmatrix3}) is of dimension $N_T \times N_T$. This should be compared to the dimensionality of the density matrix for the radiation $\widehat{\rho}(\omega,\widetilde{\omega},N_T)$, which is  $N_T! \times N_T!\;$.
Since we consider a free theory for the emitted radiation, we expect that the two-point function matrix completely determines all observables, including the density matrix. This completeness is already evident in Hawking's work \cite{info} and will
be further clarified in a future article \cite{future}.

Because the radiation matrix is Hermitian, the phase factors $e^{i\Theta_{N' N''}}$
do not appear in the expression for a physical quantity, which is obtained by tracing some operator $O$, ${\rm Tr} (\widehat{\rho}\ O)$. However, to monitor the state of the radiation,  it is essential that the phases are known rather than averaged over.

The Hawking diagonal radiation matrix is given, in terms of dimensionless
frequencies, by
\bea
\rho_H(\omega, \widetilde{\omega})
 &=& \frac{t^*_\omega t_{\wt{\omega}}}{e^{4\pi\omega}-1} \delta(\omega-\widetilde{\omega})\;,
\label{spec}
 \eea
where the $t$'s are transmission coefficients.
We assume the normalization
 $\;{\rm Tr}\; \rho_H= \int d\omega\; \rho_H(\omega,\omega)=1\;$; so that
$\;{\rm Tr}\; \rho_{SC}= \sum dN'\int d\omega\ \rho_H(\omega,\omega;N',N')=1\;$.

The frequency-dependent factor in the  off-diagonal
part of the  radiation matrix has the form
\bea
&&\Delta\rho_{OD}(\omega,\wt{\omega}) \;=\;
\frac{t^*_\omega t_{\wt{\omega}}}{(2\pi)^3} \frac{2}{(\omega \wt{\omega})^{1/2}}
\left(\frac{C_{BH}}{4}\right)^{+2i (\omega-\wt{\omega})} \cr &\times&
 \Gamma\left(1+2i\omega\right)
\Gamma(1-2i\wt{\omega})
\ \hbox{\Large\em e}^{-\pi(\omega+\wt{\omega})} \
\Gamma\left(\frac{1}{2}- i(\omega-\wt{\omega})\right)\cr
&\times&\Biggl\{
\Gamma \left(2i (\omega-\wt{\omega})\right)\left[
\frac{\Gamma \left(\frac{1}{2}+2i\wt{\omega}\right)}
{\Gamma \left(\frac{1}{2}+2i \omega\right)}
+
\frac{\Gamma \left(\frac{1}{2}-2i\omega\right)}
{\Gamma \left(\frac{1}{2}-2i \wt{\omega}\right)}
\right] + \frac{ i }{\omega-\wt{\omega}}\Biggr\}\;.
\label{rhoscf1}
\eea

The exact form  of $\Delta\rho_{OD}(\omega,\wt{\omega})$ will not be important for the rest of the discussion. It can be regarded as a uniform matrix for all dimensionless frequencies less than unity. In Schwarzschild units, this means less than the Hawking temperature $T_H$.
What will be needed in the following is
\be
{\rm Tr} \left[(\rho_{SC})^2\right]\;=\frac{1}{N_T}+\;\frac{\sqrt{2\pi}}{4}
\frac{N_{coh}(N_T)C_{BH}(N_T)}{N_T}
{\rm Tr}
\left[\left(\Delta \rho_{OD}(\omega,\wt{\omega})\right)^2\right]\;,
\label{trrho2}
\ee
where $N_{coh}$ is the coherence time in particle-number units and one finds that
\be
{\rm Tr}
\left[\left(\Delta \rho_{OD}(\omega,\wt{\omega})\right)^2\right]\;=\;
\int d \omega \int d\widetilde{\omega} \;\frac{f(\omega,\widetilde{\omega})}{\left(e^{4\pi\omega}-1\right)
\left(e^{4\pi\widetilde{\omega}}-1\right)}\;,
\ee
with $f(\omega,\widetilde{\omega})$ being  a slowly evolving function of order unity for the relevant values of frequency. It follows from the normalization convention $\;{\rm Tr}\;\rho_H=1\;$ that ${\rm Tr}\left[\left(\Delta \rho_{OD}(\omega,\wt{\omega})\right)^2\right]$ is also a number of order unity.

In the untracked case, expression~(\ref{trrho2}) is valid only until the transparency time when $N_{coh} C_{BH}\simeq 1$. As a consequence, we were unable in \cite{slowburn} to verify, quantitatively, that the radiation does  purify by the end of evaporation. The current work rectifies this earlier  omission by  using the purity and participation ratio and their relation to the Renyi entropy rather than the entanglement entropy of the radiation matrix.

Finally, let us clarify the role  of  the suppression factor
$D(N_T;N',N'')$.
Because of this factor,
the non-vanishing off-diagonal elements of $\rho_{SC}$
are contained within  a pair of strips. One of the strips extends over all permissible values of $N''$ but over a parametrically smaller
range of $N'$ values; roughly, $\;N_T-N_{coh}(N_T)\leq N' \leq N_T\;$\;.
The other strip is similar except that  $N'$ and $N''$ are interchanged.

\section{Purity and participation ratio}

\subsection{Definitions}

For a  matrix with a unit trace, the purity is defined
as $\;P[\rho]={\rm Tr}[\rho^2]\;$. For our matrix,
this is given by Eq.~(\ref{trrho2}),
\be
P(N_T)= \frac{1}{N_T}+ \frac{1}{N_T}  N_{coh}(N_T) C_{BH}(N_T)\;,
\label{purity}
\ee
where order-unity numerical factors have now been disregarded.
The first term is the contribution from the diagonal part and  will
be denoted  $P_D$, whereas  the second term is from the off-diagonal part and will be denoted  $P_{OD}$.

The participation ratio (PR), which is  defined as the inverse of the purity
$\;PR= 1/P\;$, is the main tool of our subsequent analysis. It is a measure of the number of non-vanishing eigenvalues of the radiation matrix and can,
therefore,  be expected to correspond to the Renyi entropy $H_2=\ln \left(\frac{1}{Tr ({\widehat{\rho}\;}^2)}\right)$ of the density matrix for the radiation.

Indeed, let us consider the matrix $\;{\rho}_\omega^{\ \;\widetilde{\omega}}=\langle 0_{in}|b^{\dagger}_\omega b^{\widetilde{\omega}}|0_{in}\rangle\;$, where $|0_{in}\rangle$ is the vacuum at past null infinity and $b^{\dagger}_\omega$ is a creation
operator for an outgoing Hawking mode at frequency $\omega$.  This is
essentially our radiation matrix (unnormalized, with emission times suppressed).
Then the Renyi entropy of the density matrix can be found using the simple identity \cite{future}
\be
H_2(\widehat{\rho}) = {\rm Tr}\left[\ln{(1+2{\rho})}\right]\;.
\ee
This relation makes it clear that the Renyi entropy is providing an estimate of the number of non-vanishing eigenvalues and, hence, the PR  of the matrix $\rho$. It also allows one to verify that the proposed correspondence is parametrically correct, even when the semiclassical corrections are included. This identification will be important later when our results are compared with those of the Page model.

\subsection{With tracking}

Let us first consider how to obtain the PR for our model in
the case of  ``tracking" (see Section~1). As explained, the emitted radiation has a finite coherence time $t_{coh}$ and, during this time interval, the radiation matrix elements associated with a newly  emitted Hawking mode are accessible.
Now suppose that we record the amplitude $A_{N' N''}$ and
phase $\Theta_{N' N''}$  of a  given entry at a time when this information can still be accessed.
If we do so  for all the entries, then the radiation matrix is known in its entirety at any given time. Even though some of the entries have already decayed to exponentially small, inaccessible values, we have been careful to keep records of them.

Here, we are ignoring the practical issues that are involved with tracking the information. Clearly, for recording the information coming out of a macroscopic BH, one needs a very large storage device. Also, when the radiation is tracked,
the state of the near-horizon matter changes in a significant way.
This will not be discussed here.

Next suppose that, at a given time $N_T$, we list the phases of the
measured entries and  assign to them a magnitude $\sqrt{C_{BH}(N_T)}$ rather than their originally measured magnitudes. The off-diagonal part of the radiation matrix is then a matrix of uniform magnitude $\sqrt{C_{BH}(N_T)} e^{i\Theta_{N' N''}}$.
The diagonal part is kept as it was, except that the normalization
and the temperature, which is implicit in the dimensionless frequencies,
depends on $N_T$.

In the tracked case, the coherent strips in the radiation matrix effectively extend over the
full matrix, $\;N_{coh}\to N_T\;$. Hence,
\be
P(N_T) \;=\; \frac{1}{N_T}+  C_{BH}(N_T)\;,
\label{purity1}
\ee
so that  $\;P_D=\frac{1}{N_T}\:$ and
$\;P_{OD}=C_{BH}(N_T)=\frac{1}{S_{BH}(N_T)}
\simeq\frac{1}{S_{BH}(0)-N_T}\;$.

The PR is then given by
\be
PR(N_T)\;=\; \frac{N_T(S_{BH}(0)-N_T)}{S_{BH}(0)}\;
\label{pr1}
\ee
and  starts to decrease when $\;\frac{d P}{d N_T} = 0\;$;
this being when
$\;N_T=\frac{1}{2}S_{BH}(0)\;$, exactly at the Page time! Equation~(\ref{pr1}) is depicted in Figs.~1 and 2.

For later comparison with the Page model, let us also  calculate  the deviation of the PR for our radiation matrix from that of a thermal matrix. This quantity should then  be compared to the Renyi information $\;I_R=H_2(thermal)-H_2\;$ of the density matrix; namely, the difference between the Renyi entropy of a thermal matrix and that of the actual matrix. For our model (denoting this deviation by $I_R$ as well),
\be
I_R(N_T) \;\simeq\; N_T-PR(N_T)\;=\;\frac{N_T^2}{S_{BH}(0)}\;.
\label{ourIR}
\ee
Equation~(\ref{ourIR}) is depicted in Fig.~3.

\begin{figure}
[t]
\ \hspace{1in}\scalebox{1.0} {\includegraphics{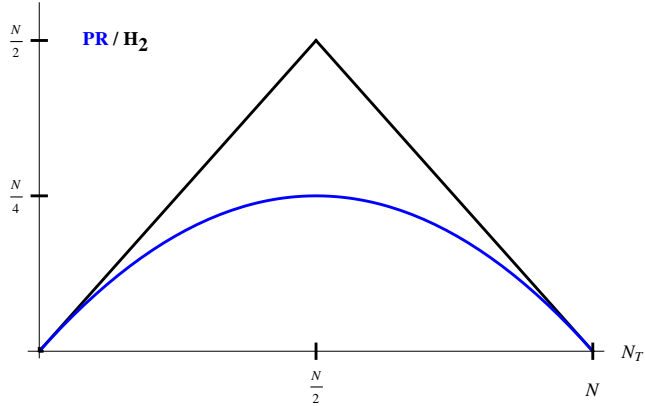}}
\caption{The Renyi entropy $H_2$ as a function of the number $N_T$ of emitted Hawking modes is shown for the Page model (black, triangular) from Eq.~(\ref{PageRenyi}) and the participation ratio  is shown for our model with tracking (blue, curved) from Eq.~(\ref{pr1}). Here, $N$ denotes the maximal value of $N_T$, which is approximately $S_{BH}(0)$.}
\end{figure}

The same results can be obtained in a way that will be relevant to the next discussion. The off-diagonal part of the purity $P_{OD}(N_T)$, when suitably normalized,  grows by one with every emission of a Hawking mode,
\be
\frac{N_T}{C_{BH}(N_T)} P_{OD}(N_T)\;=\; \frac{1}{C_{BH}(1)}P_{OD}(1) (N_T-1)\;.
\label{purity2}
\ee
Initially,  $\;P_{OD}(1)=1/S_{BH}(0)\;$. Then, taking into account that $\;C_{BH}(1) S_{BH}(0)\simeq 1\;$ and that $\;(N_T-1)/N_T \simeq 1\;$,  we have
\be
P_{OD}(N_T)\;=\; C_{BH}(N_T)\;,
\label{purity3}
\ee
exactly as in Eq.~(\ref{purity1}).

\subsection{Without tracking}

We can repeat this analysis without tracking, but first a cautionary note:  In this case, information about the correlations is only accessible after the (parametrically late) transparency time. But,  if an observer waits until after this time to record the emissions, it is unclear if she can learn about all the correlations before the  BH completely evaporates without violating the rules of causality or  seriously distorting the spacetime. The real point of this case is to show that, in principle,  the final state of the radiation purifies irrespective of the actions of any observer.

Now, in this case, the only available information during most of the lifetime of the BH is that contained within  the off-diagonal strips. The purity is, as before,
\be
P(N_T) \;=\; \frac{1}{N_T}+ \frac{1}{N_T} N_{coh}(N_T) C_{BH}(N_T)\;.
\label{Pearly}
\ee
Recall that for our model up to the transparency time,  $\;N_{coh}(N_T)=\sqrt{S_{BH}(N_T)}\;$.

We can use  Eq.~(\ref{Pearly}) to evaluate the PR. Then,  since
$\;N_{coh}(N_T)C_{BH}(N_T)=S_{BH}^{-1/2}(N_T)\ll 1\;$ and
$S_{BH}(N_T)\simeq S_{BH}(0)-N_T\;$,
\bea
PR(N_T)&=& N_T\ \frac{1}{1+  S_{BH}^{-1/2}(N_T)} \cr &\simeq& N_T\left[1-S_{BH}^{-1/2}(N_T)\right] \cr &\simeq& N_T\left[1-(S_{BH}(0)-N_T)^{-1/2}\right]\;,
\label{PRearly}
\eea
while the corresponding information  goes as
\be
I_R(N_T)\;\simeq\; N_T S_{BH}^{-1/2}(N_T)\;\simeq\;
 N_T (S_{BH}(0)-N_T)^{-1/2}\;.
\label{IRearly}
\ee
Equations~(\ref{PRearly},\ref{IRearly}) are depicted in Figs.~2 and~3, respectively.

A simple way to compare  the information content
for  untracked versus the tracked case  is to use the general form
\be
I_R(N_T) \;=\;N_T  \frac{C_{BH}(N_T)\Delta N}{1+ C_{BH}(N_T)\Delta N}\;,
\ee
where $\Delta N$ measures the degree of information accessibility;
$\;\Delta N= N_T\;$  when there is  tracking and $\;\Delta N =N_{coh}(N_T)\;$
when there is not.

The above estimates  are meant to be valid only up until the transparency time, which we will now define as the time at which the purity $P$ starts to grow. This is a somewhat different definition than that in  \cite{slowburn}, where $N_{trans}$  signaled the onset of rapid information release. However, the two definitions do give  similar results for our model.

We can formally identify $N_{trans}$ by starting with
\be
\frac{dP}{d N_T}\;=\; -\frac{1}{N_T^2}-\frac{1}{N_T^2} (S_{BH}(0)-N_T)^{-1/2}+\frac{1}{2 N_T}(S_{BH}(0)-N_T)^{-3/2}\;.
\ee
The transparency time is defined as the value of $N_T$ for which $\;\frac{dP}{d N_T}=0\;$, which occurs  when
\be
-(S_{BH}(0)-N_{trans})^{3/2}-(S_{BH}(0)-N_{trans})+\frac{1}{2} N_{trans} \;=\;0\;.
\ee
Assuming that $\;S_{BH}(0)-N_{trans} \ll S_{BH}(0)\;$ and that
$\;S_{BH}(0)-N_{trans} \gg 1\;$, we obtain
\be
N_{trans} \;=\;  S_{BH}(0)-\left(\frac{1}{2} S_{BH}(0)\right)^{{2}/{3}}\;,
\label{NTRAN}
\ee
which is parametrically equal to the transparency time as defined previously
in  \cite{slowburn}.

We would now  like to follow the purity beyond $N_{trans}$; in which case,
the previous considerations no longer apply.
However,  we can  use the fact that, after the transparency time, all the
phases and amplitudes  of the density matrix elements become available,
just as in the situation with
tracking.  This is because, by this point, the horizon no longer acts as
a causal barrier. Rather, after the transparency time, the BH acts
as a perfect reflector of information and does so
at a rate that is  only bounded by the constraints
of causality \cite{Hayden-Preskill,noburn}.

Then, as the tracking effectively starts at $N_{trans}$,
it follows by analogy with  Eq.~(\ref{purity2}) that
\be
\frac{N_T}{C_{BH}(N_T)} P_{OD}(N_T)\;=\; \frac{N_{trans}}{C_{BH}(N_{trans})} P_{OD}(N_{trans})(N_T-N_{trans})\;,
\ee
for any  $\;N_T>N_{trans}\;$.

\begin{figure}
[t]
\begin{center}
\scalebox{1.0}
{\includegraphics
{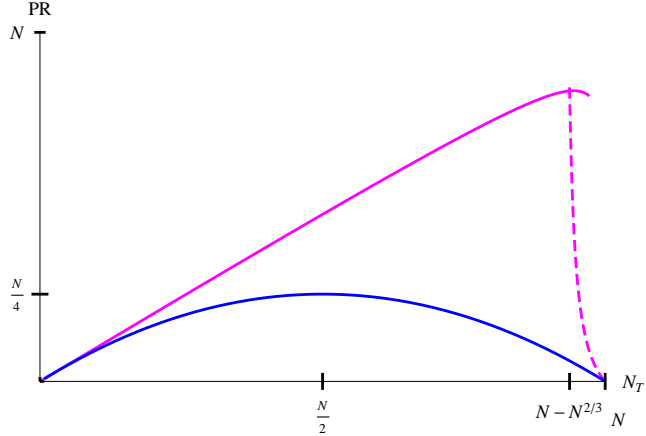}}
\end{center}
\caption{The participation ratio  as a function of the number $N_T$ of emitted Hawking modes is shown for
our model with tracking (blue, lower) from Eq.~(\ref{pr1}) and for our model without
tracking (upper solid and dashed purple). The solid purple line depicts the participation
ratio before the transparency time from Eq.~(\ref{PRearly})
and the dashed line depicts the participation ratio after the transparency time from  Eq.~(\ref{prlate}).
Here, $N$ denotes the maximal value of $N_T$, which is approximately $S_{BH}(0)$.}
\end{figure}

The above leads to
\be
 P_{OD}(N_T)\;\simeq\; \frac{1}{2}\frac{1}{S_{BH}(0)-N_T}\;,
\ee
where we have used Eqs.~(\ref{Pearly},\ref{NTRAN}) and that $N_{coh}=\sqrt{S_{BH}}$,
$\;S_{BH}(N_T)\simeq S_{BH}(0)-N_T\;$ and also
the late-time approximations  $\;N_T/N_{trans}\simeq 1\;$,
$\;N_T\simeq S_{BH}(0)\;$.

The purity reaches unity from below when
$\;N_T\to S_{BH}(0)\;$, as does the PR from above,
\be
 PR(N_T)\;=\;\frac{1}{ \frac{1}{N_T}+ \frac{1}{2}\frac{1}{S_{BH}(0)-N_T}}\;.
\label{prlate}
\ee
Equation~(\ref{prlate}) is depicted in Fig.~2.

The  information content (or deviation from thermality) approaches its maximum value of $\;S_{BH}(0)\;$
 in the same limit,
\be
I_R(N_T) \;=\; N_T\left[1-\frac{1}{ 1 + \frac{1}{2} \frac{N_T}{S_{BH}(0)-N_T}}\right]\;.
\label{IRlate}
\ee
Equation~(\ref{IRlate}) is depicted in Fig.~3.

\subsection{Comparing with Page}

For the Page model, which is cast in terms of the radiation density matrix,
 the purity was calculated by Lubkin \cite{lubkin} and later
reproduced by Page \cite{page},
\be
P_{Page}\;=\; \frac{m+n}{mn+1}\;,
\ee
where $\;m=e^{S_{rad}}\;$ and $\;n=e^{S_{BH}}\;$. And so it is given approximately (using our notations) by
\be
P_{Page}\;=\;\begin{cases}
e^{-N_T} &  N_T < S_{BH}(0)-N_T \cr
e^{-\left(S_{BH}(0)-N_T\right)} &  N_T > S_{BH}(0)-N_T\;.
\end{cases}
\ee

The corresponding Renyi entropy for the Page model is then
\be
H_{2}\;=\;\begin{cases}
N_T &  N_T < S_{BH}(0)-N_T \cr
S_{BH}(0)-N_T &  N_T > S_{BH}(0)-N_T\;,
\end{cases}
\label{PageRenyi}
\ee
while the Renyi information is
\be
I_{2}\;=\;\begin{cases}
0 &  N_T < S_{BH}(0)-N_T \cr
2 N_T- S_{BH}(0) &  N_T > S_{BH}(0)-N_T\;,
\label{IRenyiPage}
\end{cases}
\ee
which is exactly the celebrated Page curve for the von-Neumann information. Equations~(\ref{PageRenyi}), (\ref{IRenyiPage}) are depicted in Figs.~1 and 3, respectively.

Figs.~1-3 provide  graphical comparisons between our model, both with
tracking  and without, and the Page
model.

\begin{figure}
[t]
\begin{center} \scalebox{1.0} {\includegraphics{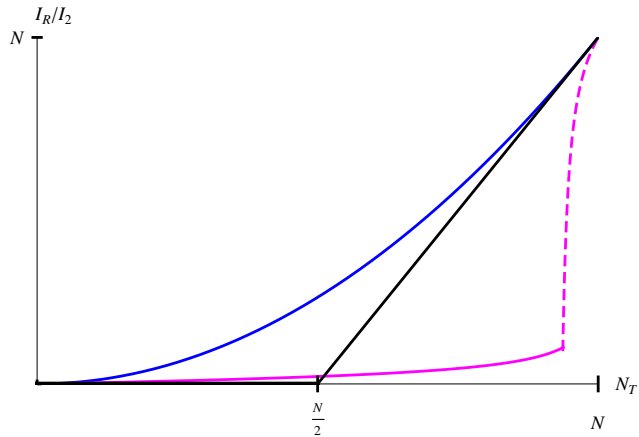}} \end{center}
\caption{The Renyi information $I_2$ as a function of the number $N_T$ of emitted Hawking modes is shown for the Page model (black) from Eq.~(\ref{IRenyiPage}) and the corresponding quantity $I_R$ is shown for our model with tracking (blue) from Eq.~(\ref{ourIR}) and without tracking (solid and dashed purple). The solid purple line depicts the information before the transparency time from Eq.~(\ref{IRearly}) and the dashed line depicts the information after the transparency time from Eq.~(\ref{IRlate}). Here, $N$ denotes the maximal value of $N_T$;
$\;N\approx S_{BH}(0)\;$.}
\end{figure}

\section{Discussion:  Pair-production perspective  and comparison of  models}

One can also view the process of Hawking radiation from the pair-production perspective. We would like to emphasize that the pair-production point of view has to be compatible with the particle-emission picture, since the ingoing modes have no effect on the fate of the outgoing particles as noted by Hawking. The pair-production picture is only relevant to the determination of the state of matter near the BH horizon.

From this viewpoint,  the pairs are initially produced in a nearly maximally entangled state. The state of the produced pairs for the untracked case was discussed in \cite{flameoff}.  We have found that the maximal entanglement remains in place for about one coherence time and then decays. We understand this behavior: After a period of one coherence time, the negative-energy modes  are absorbed into the BH and their positive-energy  partners are emitted as outgoing Hawking particles. In the tracked case, the effect of tracking changes the state of the pairs, as is standard in quantum mechanics. We intend to discuss this case in a comprehensive way in an upcoming publication \cite{schwing}.

As the negative-energy  modes  are absorbed  into the interior, they modify the  state of the BH. However, in our model, we do not keep track of the detailed state of the BH. We rather assign to the BH a wavefunction that is spherically symmetric and depends only on one parameter, the ratio of the instantaneous Schwarzschild radius to the Planck length. The wavefunction is only modified through changes to the mean value and variance of this parameter.  However, if the entries in  the off-diagonal elements of  the radiation density matrix are recorded, the detailed state of the radiation is known. Then, since the true state of the BH is the purifier of the radiation, this is equivalent to knowing
the details about  the BH state. One  might view this  as a version of BH complementarity.

Meanwhile, in the Page model, the interior contents of the BH are entirely traced over, and so nothing can be known  about the initial state of the BH until after  the Page time has transpired. This is because, in Page's construction,  most of the information is stored as correlations within whatever is the larger of the two subsystems, the radiation or the BH interior \cite{page}. This is also the reason why the first bit of information only comes out after the Page time, whereas the first bit emerges after the parametrically shorter time interval $t_{coh}$  in  our framework \cite{slowleak}.

The same observation also underlies why the Page model represents the minimal
rate at which the radiation from a burning body can be purified: One is tracing out the maximal amount of information   at all times in the process. Conversely,
our model with tracking requires a  minimal amount of tracing; the negative-energy partners  and a single-parameter wavefunction. Therefore, this corresponds
to  a higher rate of purification. We will go one step further and conjecture
that our model corresponds to the {\em highest} possible rate
of purification.

\section*{Acknowledgments}
We would like to thank Lasma Alberte and Andrei Khmelnitsky for many useful discussions and  suggestions and for comments on the  manuscript.  We also thank Gary Gibbons, Sunny Itzhaki, Samir Mathur, Don Page and David Turton for discussions.  The research of RB was supported by the Israel Science Foundation grant no. 239/10. The research of AJMM received support from an NRF Incentive Funding Grant 85353 and a Rhodes Discretionary Grant RD31/2013. AJMM thanks Ben Gurion University for their  hospitality during his visit.

\end{document}